\documentclass[12pt]{article}

\usepackage{epsfig}

\newcommand{\be}{\begin{eqnarray}}
\newcommand{\ee}{\end{eqnarray}}

\def\lsim{\mathrel{\rlap{\lower4pt\hbox{\hskip1pt$\sim$}}
    \raise1pt\hbox{$<$}}}               
\def\gsim{\mathrel{\rlap{\lower4pt\hbox{\hskip1pt$\sim$}}
    \raise1pt\hbox{$>$}}}               

\newcommand{\<}{\langle}
\renewcommand{\>}{\rangle}
\newcommand{\GGt}{ {G \widetilde{G}} }
\newcommand{\de}{\partial}
\newcommand{\mc}{\mathcal}

\begin{document}

\rightline{\Large Preprint RM3-TH/02-16}
\rightline{\Large ROMA-1343-02}

\vspace{1cm}

\begin{center}

\LARGE{Neutron Electric Dipole Moment on the Lattice:\\[2mm] a Theoretical Reappraisal}

\vspace{1cm}

\large{D. Guadagnoli$^1$, V. Lubicz$^{2,3}$, G. Martinelli$^{1,4}$ and S. Simula$^3$}

\vspace{0.5cm}

\normalsize{$^1$Dipartimento di Fisica, Universit\`a di Roma ``La Sapienza"\\ Piazzale Aldo Moro 2, I-00185 Roma, Italy\\ $^2$Dipartimento di Fisica, Universit\`a di Roma Tre\\ Via della Vasca Navale 84, I-00146 Roma, Italy\\ $^3$Istituto Nazionale di Fisica Nucleare, Sezione Roma III\\ Via della Vasca Navale 84, I-00146 Roma, Italy\\ $^4$Istituto Nazionale di Fisica Nucleare, Sezione di Roma\\ Piazzale Aldo Moro 2, I-00185 Roma, Italy}

\end{center}

\vspace{1cm}

\begin{abstract}

\noindent We present a strategy for a lattice evaluation of the neutron electric dipole moment induced by the strong $CP$ violating term of the $QCD$ Lagrangian. Our strategy is based on the standard definition of the electric dipole moment, involving the charge density operator $J_0$, in case of three flavors with non-degenerate masses. We present a complete diagrammatic analysis showing how the axial chiral Ward identities can be used to replace the topological charge operator with the flavor-singlet pseudoscalar density $P_S$. Our final result is characterized only by disconnected diagrams, where the disconnected part can be either the single insertion of $P_S$ or the separate insertions of both $P_S$ and $J_0$. The applicability of our strategy to the case of lattice formulations that explicitly break chiral symmetry, like the Wilson and Clover actions, is discussed.

\end{abstract}

\vspace{1cm}

PACS numbers: 11.30.Er, 13.40.Ern, 14.20.Dh, 12.38.Gc

\vspace{0.25cm}

Keywords: \parbox[t]{12cm}{Time reversal symmetry; Electric dipole moment of the neutron; Lattice QCD.}

\newpage

\pagestyle{plain}

\section{Introduction \label{section:introduction}}

\indent A non-vanishing value of the electric dipole moment ($EDM$) of the neutron is a direct manifestation of the breakdown of both parity and time reversal symmetries. Within the Standard Model and its possible extensions parity and time reversal can be violated both in the electroweak and strong sectors. In this study we focus on the neutron $EDM$ induced by the so-called $\theta$-term of the $QCD$ Lagrangian~\cite{tHooft}, given in the Euclidean space by
 \be
    {\cal{L}}_{\theta} = i \theta {g^2 \over 32 \pi^2} \GGt = i\theta {g^2 
    \over 64 \pi^2} \varepsilon_{\alpha \beta \mu \nu} G_{\alpha \beta}^c
    G_{\mu \nu}^c ~ ,
    \label{eq:theta}
 \ee
where $G_{\mu \nu}^c$ is the gluon field strength, $c$ the color octet index ($c = 1, ..., 8$) and $\theta$ a dimensionless parameter.

\indent The present experimental upper limit on the neutron $EDM$, $d_N \equiv |\vec{d}_N| < 6.3 \cdot 10^{-26} ~ (e \cdot cm)$ at $90 \%$ confidence level~\cite{ILL}, which corresponds to a severe bound on the magnitude of $\theta$. Indeed, using the available theoretical estimates from Refs.~\cite{Baluni,Crewther}, one has $d_N \approx 3 \cdot 10^{-16} ~ |\theta| ~ (e \cdot cm)$ leading to $|\theta| \lsim 2 \cdot 10^{-10}$. The smallness of the parameter $\theta$ is usually referred to as the strong $CP$ problem.

\begin{figure}[t]

\centerline{\epsfig{file=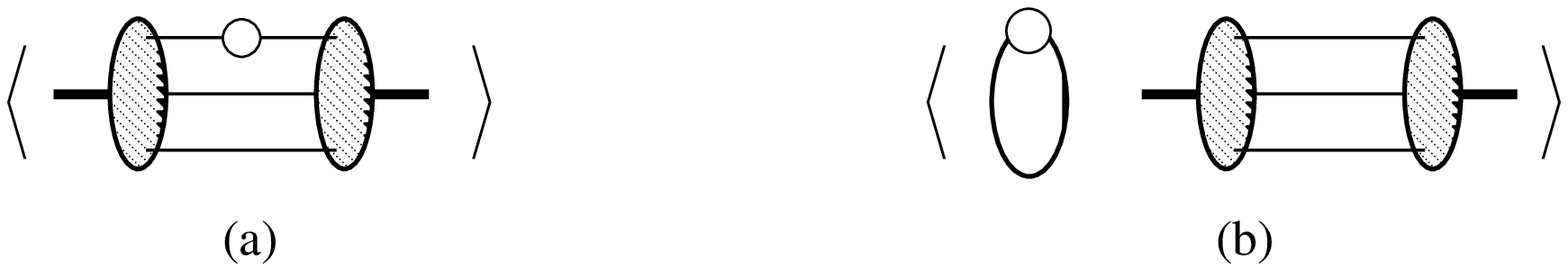,width=6.75in}}

\caption{\em Connected (a) and disconnected (b) insertions of the flavor-singlet pseudoscalar density operator. The insertions are represented by open dots, while solid lines are quark propagators. The hatched ovals are the operators which create and destroy the neutron. Gluon lines as well as extra quark loops are not shown.}

\label{fig:diag1}

\end{figure}

\indent Available estimates of the relevant matrix element however are based on phenomenological models, as the $MIT$ bag model of Ref.~\cite{Baluni} or as the effective $\pi N$ chiral Lagrangian of Ref.~\cite{Crewther}. Estimates relying on non-perturbative methods based on the fundamental theory, like lattice $QCD$, are still missing to date. A first attempt to calculate $d_N$ with lattice $QCD$ was done in Ref.~\cite{AG89} long time ago. There, using the axial anomaly (as suggested in Ref. ~\cite{Baluni}), the $\theta$-term (\ref{eq:theta}) was replaced by the flavor-singlet pseudoscalar density $P_S$, obtaining
 \be
    {\cal{L'}}_{\theta} = - {i \over N_f} \theta ~ \overline{m} P_S 
    \rightarrow - {i \over 3} \theta ~ \overline{m} \left[ \bar{u} \gamma_5 
    u + \bar{d} \gamma_5 d + \bar{s} \gamma_5 s \right] ~ ,
    \label{eq:P5}
 \ee
where $\overline{m}^{-1} \equiv (m_u^{-1} + m_d^{-1} + m_s^{-1}) / 3$ in case of three flavors with non-degenerate masses. Then, $d_N$ was evaluated by extracting the spin-up and spin-down neutron masses from the two-point functions obtained by adding to the action both ${\cal{L'}}_{\theta}$ and a term corresponding to a constant-background electric field oriented in a fixed spatial direction. Since $\theta$ is expected to be quite small, it is enough to consider a single insertion of ${\cal{L'}}_{\theta}$ in the two-point functions. The latter can be divided into connected and disconnected insertions, when the operator (\ref{eq:P5}) is contracted in a valence-quark line and in a virtual quark loop, respectively, as pictorially represented in  Fig.~\ref{fig:diag1}.

\indent In Ref.~\cite{AG89} a lattice calculation using Wilson fermions and including only connected diagrams was attempted and a non-vanishing result for $d_N$ was found. In Ref.~\cite{Aoki} it was shown however that the connected diagrams cannot contribute to the neutron $EDM$ and therefore the non-zero result of Ref.~\cite{AG89} arises from lattice artifacts of $O(a)$. Thus a lattice $QCD$ estimate of the neutron $EDM$ is not yet available.

\indent It should be pointed out that the approach of Refs.~\cite{AG89,Aoki} is based on the technique of introducing an external constant-background field. This approach has been abandoned since modern lattice $QCD$ simulations allow to evaluate directly the relevant matrix elements of the electromagnetic ($e.m.$) current operator. Moreover, in the proof of Ref.~\cite{Aoki} it is assumed that the connected insertions of the singlet and non-singlet pseudoscalar densities are the same. Since this is not the case as we have explicitly checked, a more careful diagrammatic analysis of the insertions of the pseudoscalar operators appears to be worthwhile.

\indent In this Letter we present an alternative way to compute the neutron $EDM$ on the lattice, which avoids the introduction of the background electric field thanks to the use of the standard definition of the neutron $EDM$
 \be
    \vec{d}_N \equiv \int d^3y ~ \vec{y} ~ _{\theta}\langle N | J_0(y) | N
    \rangle_{\theta} ~ , 
    \label{eq:dipole}
 \ee
where $J_0$ is the time component of the $e.m.$ current operator. If the $\theta$-term (\ref{eq:theta}) is neglected in the $QCD$ Lagrangian, then the moment $\vec{d}_N$ identically vanishes. Treating ${\cal{L}}_{\theta}$ as a perturbation at first order, one has
 \be
    \vec{d}_N \equiv -i \theta {g^2 \over 32 \pi^2} \int d^3y ~ \vec{y} ~
    _0\langle N | J_0(y) \left[ \int d^4x ~ \GGt(x) \right] | N \rangle_0 ~ 
    , 
    \label{eq:EDM}
 \ee
where $| N \rangle_0$ is a shorthand for $| N \rangle_{\theta = 0}$.

\indent As well known~\cite{Gockeler}, the direct calculation of Eq.~(\ref{eq:EDM}) is a difficult task due to the presence of the topological charge operator. Therefore it is of interest to investigate the possibility of replacing the insertion of the topological charge with the disconnected insertions of the flavor-singlet pseudoscalar density in the presence of the charge density operator $J_0$. Since the insertions of the latter can generate disconnected diagrams, our analysis represents also an extension of the previous one made in Ref.~\cite{Aoki}.

\indent The plan of the paper is as follows. In Section~\ref{section:Ward} we briefly recall the non-singlet and singlet continuum axial Ward Identities ($WI$'s) which are relevant in this work. In Section~\ref{section:result} we present a complete diagrammatic analysis, showing that the topological charge can be replaced by the flavor-singlet pseudoscalar density in Eq.~(\ref{eq:EDM}). Our final result is characterized only by disconnected diagrams, where the disconnected part can be either the single insertion of $P_S$ or the separate insertions of both $P_S$ and $J_0$. The applicability of our strategy to the case of lattice formulations that explicitly break chiral symmetry, like the Wilson and Clover actions, is discussed. Our conclusions are summarized in Section~\ref{section:conclusions}.

\section{Axial Ward identities \label{section:Ward}}

\indent Axial $WI$'s can be obtained in the usual way starting from the vacuum expectation value of a generic operator $\mathcal{O}$. In Euclidean space one has
 \be
    \< \mc{O}(x_1,...,x_n) \> = {1 \over Z_0} \int d[G] d[\psi] 
    d[\bar{\psi}] ~ \mc{O}(x_1,...,x_n) ~ e^{-S}
    \label{eq:vev}
 \ee
where $S$ is the $QCD$ action without the $\theta$-term and $Z_0 \equiv \int d[G] d[\psi] d[\bar{\psi}] ~ e^{-S}$. From now on the notation $\langle ... \rangle$ refers to vacuum expectation values evaluated without the $\theta$-term (\ref{eq:theta}) in the action. Performing the non-singlet axial rotations of the quark fields $\psi(x)$ and $\bar{\psi}(x)$
\be
   & & \psi(x) \rightarrow \left[ 1 + i \alpha^a(x) \lambda^a \gamma_5 
   \right] \psi(x) ~ , \nonumber \\[2mm]
   & & \bar{\psi}(x) \rightarrow \bar{\psi}(x) \left[ 1 + i \alpha^a(x) 
   \lambda^a \gamma_5 \right] ~ ,
   \label{eq:NS}
 \ee
where $\lambda^a$ are the usual $SU(3)$ flavor matrices ($a = 1, ..., 8$), one gets
 \be 
    \< \mathcal{O} {\delta S \over \delta(i \alpha^a(x))} \> = \< {\delta 
    \mathcal{O} \over \delta(i \alpha^a(x))} \> 
    \label{eq:NSWIa}
 \ee
with
 \be 
    \< \mathcal{O} {\delta S \over \delta(i \alpha^a(x))} \> = - \de_{\mu} 
    \< \mc{O} A_{\mu}^a(x) \> + \< \mc{O} ~ \bar{\psi}(x) \left\{ M, 
    {\lambda^a \over 2} \right\} \gamma_5 ~ \psi(x) ~ \> ~ .
    \label{eq:NSWIb}
 \ee
where $M$ is the $SU(3)$ mass matrix and $A_{\mu}^a(x)$ the non-singlet axial current. From now on all the relevant operators and fields should be considered as renormalized quantities in the continuum. The case of the lattice regularization will be addressed at the end of Section~\ref{section:result}.

\indent Performing the flavor-singlet rotations
 \be
    \psi(x) & \rightarrow & \left[ 1 + i \alpha^0(x) \gamma_5 \right] 
    \psi(x) ~ , \nonumber \\[2mm]
    \bar{\psi}(x) & \rightarrow & \bar{\psi}(x) \left[ 1 + i \alpha^0(x) 
    \gamma_5 \right] ~ ,
    \label{eq:S}
 \ee
one gets
 \be
    \< \mathcal{O} {\delta S \over \delta(i \alpha^0(x))} \> = \< {\delta 
     \mathcal{O} \over \delta(i \alpha^0(x))} \> 
    \label{eq:SWIa}
 \ee
where
 \be
    \< \mathcal{O} {\delta S \over \delta(i \alpha^0(x))} \> = - \de_{\mu} 
    \<\mc{O} A_{\mu}^S(x) \> + \< \mc{O} ~ \bar{\psi}(x) \left\{ M, 
    \lambda_0 \right\} \gamma_5 ~ \psi(x) ~ \> + 2 N_f {g^2 \over 32 \pi^2} 
    \< \mc{O} \GGt(x) \> ~ ,
    \label{eq:SWIb}
\ee
with $\lambda_0$ being the $SU(3)$ identity matrix and $A_{\mu}^S(x)$ the (renormalized) flavor-singlet axial current. In Eq.~(\ref{eq:SWIb}) the anomalous term proportional to the (renormalized) $\GGt$ operator has been taken into account and $g$ is the renormalized strong coupling. Note that in the non-singlet channel all the terms appearing in the r.h.s of Eqs.~(\ref{eq:NSWIb}) are separately renormalization-group invariant, while in the singlet case [Eq.~(\ref{eq:SWIb})] a divergent mixing between $\de_{\mu} A_{\mu}^S$ and $g^2 ~ \GGt$ appears.

\indent Following the trick suggested in Ref.~\cite{Baluni}, Eqs.~(\ref{eq:SWIa}) and (\ref{eq:SWIb}) can be used to replace $\GGt(x)$ in Eq.~(\ref{eq:EDM}) with the pseudoscalar density $P_S(x)$. By integrating the $WI$ over the whole space-time, the divergence of the axial current $\de_{\mu} A_{\mu}^S$, appearing in the r.h.s. of Eq.~(\ref{eq:SWIb}), vanishes identically\footnote{This is true in absence of massless Goldstone-boson poles, which certainly occurs away from the chiral limit. In case of the singlet channel the space-time integration of $\de_{\mu} A_{\mu}^S$ vanishes also in the chiral limit of full $QCD$, when the $\eta'$-meson remains massive because of the gluon anomaly. Let us remind however that in the chiral limit the $\theta$-term (\ref{eq:theta}) can be eliminated by a chiral rotation of the quark fields and therefore in that limit $d_N$ must vanish.}, whereas the flavor-singlet contact terms given by the variation of $\mc{O}$ in the r.h.s. of Eq.~(\ref{eq:SWIa}), may give a non-null contribution.

\indent Without loss of generality we can limit ourselves to the simple $SU(2)$-symmetric case, where $m_u = m_d = m$, and $m_s \neq m$. Since
 \be
    \left\{ M, \lambda_0 \right\} = 2m \left[ \left(1 + {\eta_s \over 3} 
    \right) \lambda_0 - {\eta_s \over \sqrt{3}} \lambda^8 \right]
    \label{eq:Mlambda0}
 \ee
where $\eta_s = (m_s - m) / m$, the singlet $WI$ (\ref{eq:SWIb}) involves both the flavor-singlet pseudoscalar density $P_S$ and the non-singlet operator $P^8$. To eliminate the latter we have to combine the singlet $WI$ (\ref{eq:SWIb}) and the non-singlet $WI$ (\ref{eq:NSWIb}) with $a = 8$. By exploiting the relation
 \be
    \left\{ M, {\lambda^8 \over 2} \right\} = 2m \left[ \left(1 + {2 \eta_s 
    \over 3} \right) {\lambda^8 \over 2} - {\eta_s \over 3 \sqrt{3}} 
    \lambda_0 \right] ~ ,
    \label{eq:Mlambda8}
 \ee
one finds
 \be
    \left\{ M, \lambda_0 \right\} + {2 \eta_s \over \sqrt{3}} {1 \over 1 + 2 
    \eta_s / 3} \left\{ M, {\lambda^8 \over 2} \right\} = 2 \overline{m} 
    \lambda_0 ~ ,
    \label{eq:mbar}
 \ee
with $\overline{m}^{-1} = (m^{-1} + m^{-1} + m_s^{-1}) / 3$. Equation~(\ref{eq:mbar}) implies
 \be
     2 N_f {g^2 \over 32 \pi^2} \int d^4x ~ \< \mc{O} \GGt(x) \> =  \< 
     \delta[\mc{O}] \> - 2 \overline{m} ~ \int d^4x ~ \< \mc{O} P_S(x) \> 
     \label{eq:master}
 \ee
with
 \be
     \< \delta[\mc{O}] \> = \< \delta_S[\mc{O}] \> + {2 \eta_s \over 
     \sqrt{3}} {1 \over 1 + 2 \eta_s / 3} \< \delta_8[\mc{O}] \>
     \label{eq:delta} ~ ,
 \ee
where $\delta_S[\mc{O}]$ and $\delta_8[\mc{O}]$ stand for the variations of $\mc{O}$ integrated over the whole space-time and corresponding to the singlet and octet chiral rotations [see Eqs. (\ref{eq:S}) and (\ref{eq:NS})], respectively. In the next Section we present a complete diagrammatic analysis of the r.h.s. of Eq.~(\ref{eq:master}).

\section{Neutron $EDM$ and the disconnected insertions of the pseudoscalar density \label{section:result}}

\indent In the flavor-singlet channel the operator $\mc{O}$ of interest for Eq.~(\ref{eq:EDM}) is given by
 \be
    \mc{O} = N_{\alpha}(z) ~ J_0(y) ~ \bar{N}_{\beta}(0)
    \label{eq:O_S}
 \ee
where $J_0(y)$ is the $e.m.$ charge density
 \be
    J_0(y) = e_u ~ \bar{u}(y) \gamma_0 u(y) +e_d ~ \bar{d}(y) \gamma_0 d(y) 
    + e_s ~ \bar{s}(y) \gamma_0 s(y) ~ ,
    \label{eq:J0}
 \ee
and $N_{\alpha (\beta)}$ are the interpolating fields of the neutron~\cite{Ioffe}
 \be
    N_{\alpha} = {1 \over \sqrt{6}} \varepsilon^{i j k} d_{\alpha}^i \left( 
    d^{T j} \mc{C} \gamma_5 u^k \right)
    \label{eq:neutron}
 \ee
with $i, j, k$ being color indexes, $\alpha$ and $\beta$ Dirac spinor indexes, and $\mc{C}$ the charge conjugation operator. As well known, standard procedures in lattice $QCD$ calculations allow to extract the relevant matrix element using the interpolating fields (\ref{eq:neutron}) at enough large values of the source (sink) time distances.

\begin{figure}[t]

\centerline{\epsfxsize=6.75in \epsfbox{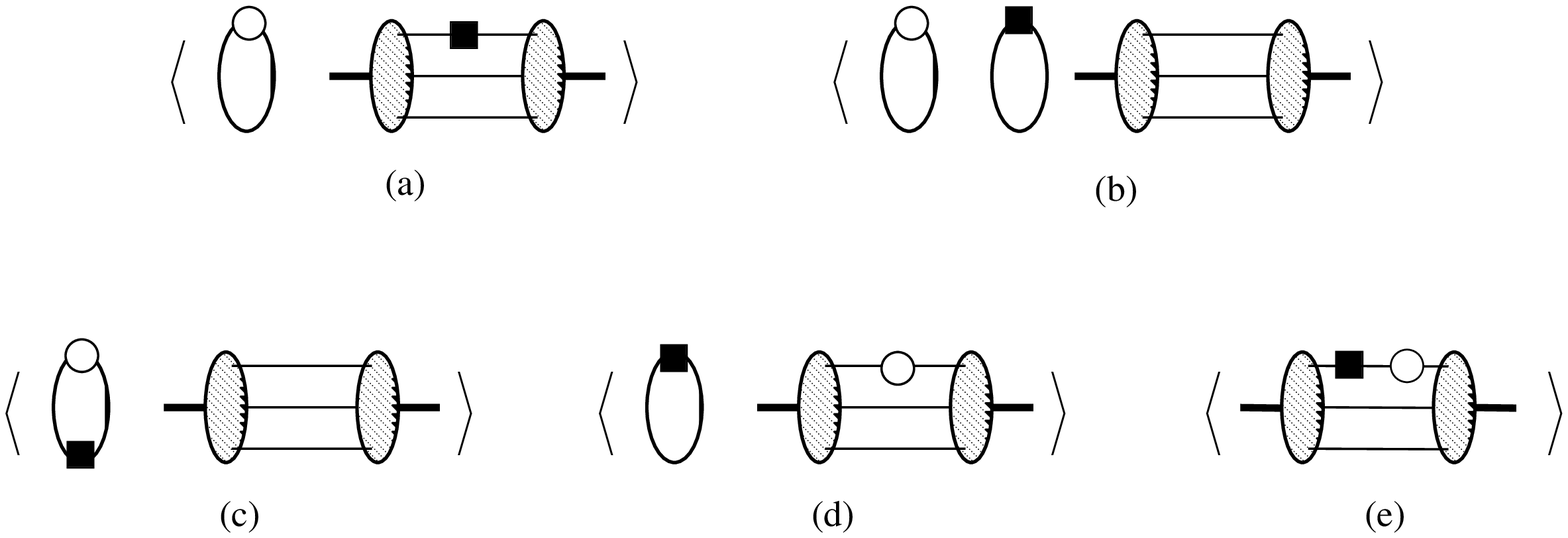}}

\caption{\em Possible typologies of disconnected diagrams [$(a)$ - $(d)$] and an example of a connected diagram [$(e)$], involved in $\mc{O} P_S(x)$. The meaning of the vertices and lines is the same as in Fig.~\ref{fig:diag1}. Open dots and full squares represent the insertion of $\gamma_5$ ($P_S$) and $\gamma_0$ ($J_0$), respectively.}

\label{fig:diag2}

\end{figure}

\begin{figure}[htb]

\centerline{\epsfig{file=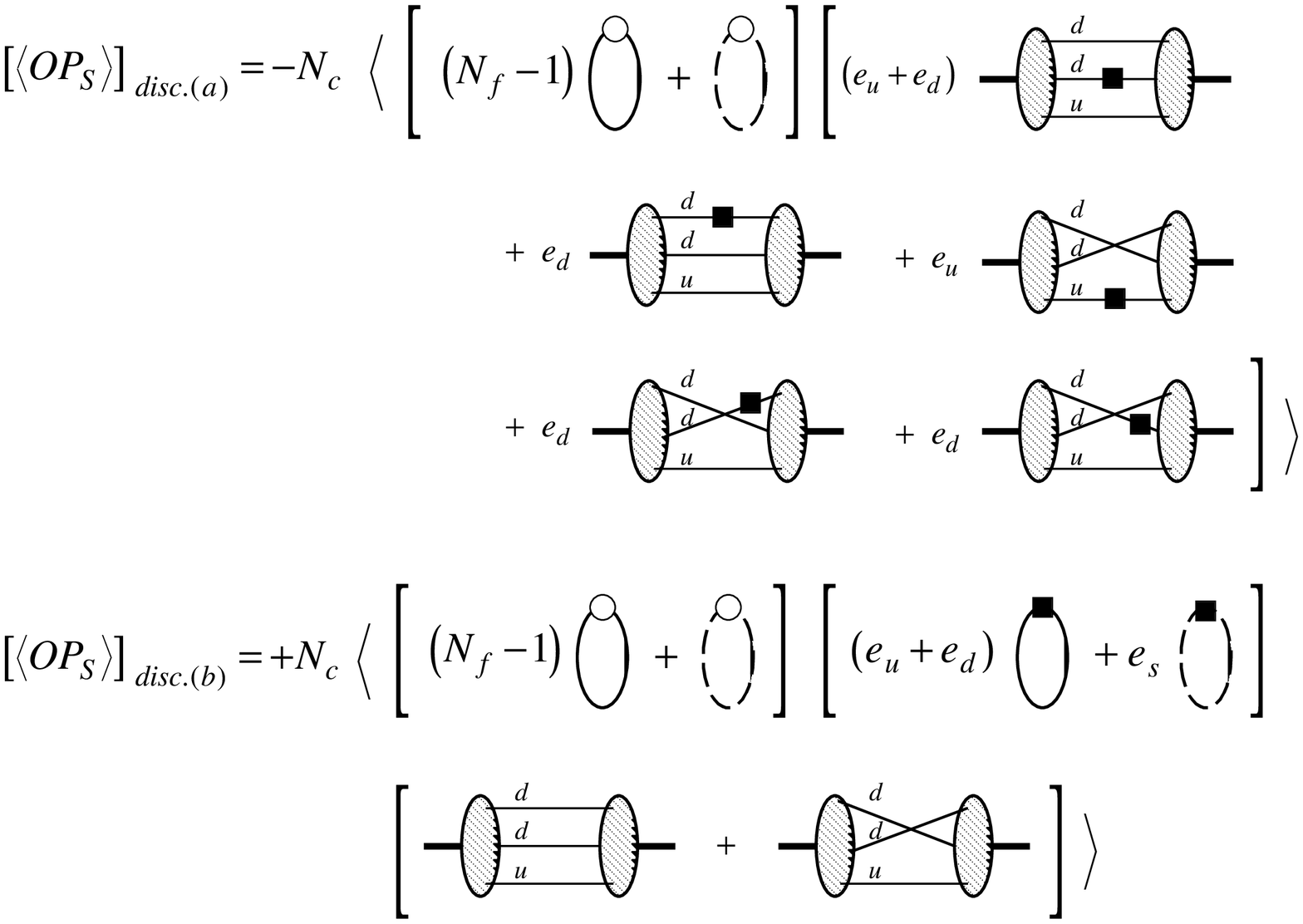,width=6.75in}}

\caption{\em Disconnected diagrams contributing to the r.h.s. of Eq.~(\ref{eq:result}). The notation is the same as in Fig.~\ref{fig:diag2}, but solid lines are now $u$- and $d$-quark propagators, while dashed lines are $s$-quark propagators.}

\label{fig:diag3}

\end{figure}

\indent Let us now consider the chiral variations $\delta_S[\mc{O}]$ and $\delta_8[\mc{O}]$ of the operator (\ref{eq:O_S}), which can be concisely written in the form 
 \be
    \delta_j[\mc{O}] = \delta_j[N_{\alpha}](z) ~ J_0(y) ~ \bar{N}_{\beta}(0) 
    + N_{\alpha}(z) ~ \delta_j[J_0](y) ~ \bar{N}_{\beta}(0) + N_{\alpha}(z) 
    ~ J_0(y) ~ \delta_j[\bar{N}_{\beta}](0)
    \label{eq:deltaO}
 \ee 
where $j$ stands for $j = S$ or $j = 8$. Through the package $FORM$~\cite{FORM} we have carried out all the Wick contractions involved in the variations (\ref{eq:deltaO}) as well as in the quantity $\mc{O} P_S(x)$. The Feynman diagrams corresponding to the chiral variations $\delta_S[\mc{O}]$ and $\delta_8[\mc{O}]$ cannot be directly compared with those corresponding to $\mc{O} P_S(x)$, because the former contain one quark propagator less than the latter. In order to make the comparison possible we make use of the non-singlet axial $WI$'s (\ref{eq:NSWIa}) and (\ref{eq:NSWIb}) with the operator $\mc{O}$ given by $\mc{O} = \psi(w) ~ \bar{\psi}(w')$. After integration over the whole space-time one obtains
 \be
     S_i(w, w') ~ \gamma_5 + \gamma_5 ~ S_i(w, w') = 2 m_i ~ \int d^4x ~ 
     S_i(w, x) ~ \gamma_5 ~ S_i(x, w')
     \label{eq:motion}
 \ee
where $S_i$ is the propagator of the $i$-th quark. The systematic use of Eq.~(\ref{eq:motion}) in the Feynman diagrams generated by the chiral variations of the operator $\mc{O}$ allows to gain one more propagator and one more integration in a such a way that the diagrams arising from the chiral variations can be now directly compared with the Feynman diagrams generated by the operator $\mc{O} P_S(x)$. The latter can be divided into connected and disconnected diagrams, whose possible typologies are sketched in Fig.~\ref{fig:diag2}. The chiral variations cancel out all the connected diagrams [see Fig.~\ref{fig:diag2}(e)], and therefore we confirm the final result of Ref.~\cite{Aoki} that connected diagrams do not contribute to the neutron $EDM$. However, the chiral variations contain also disconnected diagrams of the type $(c)$ and $(d)$ which cancel out the corresponding diagrams present in $\mc{O} P_S(x)$. While diagrams of the type $(c)$ vanish identically, as it can be checked using Eq.~(\ref{eq:motion}) and the relation $\{ \gamma_5, \gamma_0 \} = 0$, the diagrams of the type $(d)$ are non-vanishing in case of $SU(3)$-symmetry breaking. Therefore, our general result is
 \be
    2 N_f {g^2 \over 32 \pi^2} \int d^4x ~ \< \mc{O} \GGt(x) \> = - 2 
    \overline{m} \int d^4x ~ \left\{ ~ \left[ \< \mc{O} P_S(x) \> 
    \right]_{disc. (a)} + \left[ \< \mc{O} P_S(x) \> \right]_{disc. (b)} ~ 
    \right\} ~ ,
    \label{eq:result}
 \ee
which allows to replace in Eq.~(\ref{eq:EDM}) the topological charge with disconnected insertions of the pseudoscalar density. A pictorial representation of all the explicit diagrams contributing to $\left[ \< \mc{O} P_S(x) \> \right]_{disc. (a)}$ and $\left[ \< \mc{O} P_S(x) \> \right]_{disc. (b)}$ is shown in Fig.~\ref{fig:diag3}.

\indent Note that: ~ i) in the $SU(3)$ symmetric limit the diagrams of type $(d)$ vanish, because the disconnected insertions of the charge density $J_0$ are zero due to the relation $e_u + e_d + e_s = 0$; ~ ii) the r.h.s. of Eq.~(\ref{eq:result}) is proportional to $\overline{m}$. Since $\overline{m} = 0$ when at least one of the quark masses is zero, the insertion of the topological charge has no effect in such a limit and the neutron $EDM$ is vanishing; ~ iii) the only insertions of the singlet pseudoscalar density, which are left in the final result (\ref{eq:result}), are those disconnected diagrams which are related to the operator $\GGt$ via the anomaly. This is why the diagrams of the types $(c)$ and $(d)$ of Fig.~\ref{fig:diag2} do not contribute to the neutron $EDM$.

\begin{figure}[htb]

\centerline{\epsfig{file=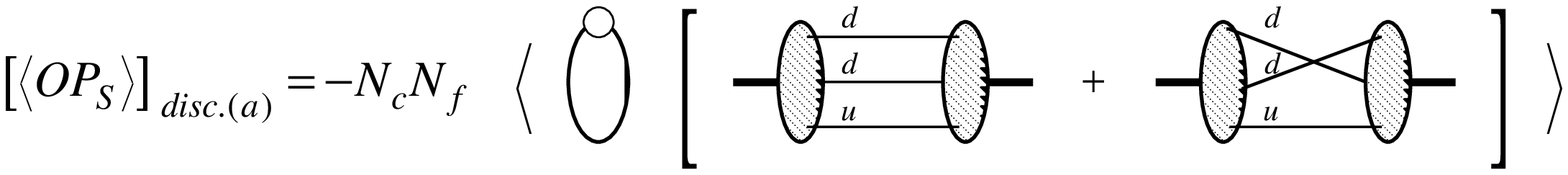,width=6.75in}}

\caption{\em Disconnected insertions of the flavor-singlet pseudoscalar density when the $SU(3)$ symmetric limit is considered and the operator $\mc{O}$ [Eq.~(\ref{eq:O_S})] does not contain the charge density operator $J_0$. The meaning of the vertices and lines is the same as in Fig.~\ref{fig:diag3}.}

\label{fig:diag4}

\end{figure}

\indent The final result of Ref.~\cite{Aoki} can be recovered from Eq.~(\ref{eq:result}) simply by considering the $SU(3)$ symmetric limit and by dropping the charge density $J_0$ in the operator $\mc{O}$ [see Eq.~(\ref{eq:O_S})]. In this case $\left[ \< \mc{O} P_S(x) \> \right]_{disc. (b)} = 0$ and the disconnected contributions $\left[ \< \mc{O} P_S(x) \> \right]_{disc. (a)}$ are those illustrated in Fig.~\ref{fig:diag4}.

\indent Before closing this Section, we show the applicability of our results to the case of lattice formulations, that break explicitly chiral symmetry, as the Wilson or Clover actions. To do that we consider the works of Refs.~\cite{KS,Guido}, whose central result is that even within the above-mentioned formulations the continuum limit can be reached in such a way that the continuum $WI$'s are respected.

\indent When the functional integral is regularized on the lattice using Wilson fermions~\cite{Wilson}, the effect of a non-singlet axial rotation leads to the following identity:
 \be
    - \Delta_{\mu} \< \mc{O} \hat{A}_{\mu}^a(x) \> + \< \mc{O} ~ 
    \hat{\bar{\psi}}(x) \left\{ M_0, {\lambda^a \over 2} \right\} \gamma_5 ~ 
    \hat{\psi}(x) \> + \< \mc{O} X^a(x) \> = \< {\delta[\mc{O}] \over 
    \delta(i\alpha^a(x))} \>
    \label{eq:WIa}
 \ee
where $\hat{A}_{\mu}^a$ and $\hat{\psi}$ are a lattice version of the (non-singlet) bare axial current and quark field, respectively, and $M_0$ is the bare $SU(3)$ quark mass matrix. In Eq.~(\ref{eq:WIa}) $\Delta_{\mu}$ is the backward derivative on the lattice, while the operator $X^a(x)$ is the chiral variation of the Wilson term. The latter is known~\cite{Guido} to mix with operators of the same and lower dimensionality, namely
 \be
    \< \mc{O} X^a(x) \> =  \left[1 - Z_A^{(NS)} \right] \Delta_{\mu} \< 
    \mc{O} \hat{A}_{\mu}^a(x) \> - \< \mc{O} ~ \hat{\bar{\psi}}(x) \left\{ 
    \overline{M}^{(NS)}, {\lambda^a \over 2} \right\} \gamma_5 ~ 
    \hat{\psi}(x) \> - \< \mc{O} \overline{X}^a(x) \>
    \label{eq:Xa}
 \ee
where $Z_A^{(NS)}$ and $\overline{M}^{(NS)}$ are mixing coefficients, and $\overline{X}^a(x)$ is an operator whose matrix elements vanish in the continuum limit. Substituting Eq.~(\ref{eq:Xa}) into Eq.~(\ref{eq:WIa}) one gets
 \be
    - \Delta_{\mu} \< \mc{O} \left[ Z_A^{(NS)} \hat{A}_{\mu}^a(x) \right] \> 
    + \< \mc{O} \left[ Z_P^{(NS)} \hat{\bar{\psi}}(x) \left\{ M^{(NS)},
    {\lambda^a \over 2} \right\} \gamma_5 ~ \hat{\psi}(x) \> \right] \> = 
    \< {\delta[\mc{O}] \over \delta(i\alpha^a(x))} \> + O(a)
    \label{eq:preWIa}
 \ee
where $M^{(NS)} \equiv [M_0 - \overline{M}^{(NS)}] / Z_P^{(NS)}$ with $Z_P^{(NS)}$ being the renormalization constant of the non-singlet pseudoscalar density. Therefore, one can identify the continuum limit of $Z_A^{(NS)} \hat{A}_{\mu}^a(x)$ and $Z_P^{(NS)} \hat{\bar{\psi}}(x) \left\{ M^{(NS)}, \lambda^a / 2 \right\} \gamma_5 ~ \hat{\psi}(x)$ with the renormalized axial current $A_{\mu}^a(x)$ and $\bar{\psi}(x) \left\{ M^{(NS)}, \lambda^a / 2 \right\} \gamma_5 ~ \psi(x)$, respectively, appearing in Eq.~(\ref{eq:NSWIb}). In this way the non-singlet continuum $WI$ is recovered with a renormalized quark mass matrix given by $M^{(NS)}$.

\indent In the singlet case the chiral variation of the Wilson term, $X^S$, obeys the relation
 \be
    \< \mc{O} X^S(x) \> & = & \left[1 - Z_A^{(S)} \right] \Delta_{\mu} \< 
    \mc{O} \hat{A}_{\mu}^S(x) \> + counterterms - \< \mc{O} ~ 
    \hat{\bar{\psi}}(x) \left\{ \overline{M}^{(S)}, \lambda_0 \right\} 
    \gamma_5 ~ \hat{\psi}(x) \> \nonumber \\
    & + & { 2 N_f \over 32 \pi^2} Z_{(g_0^2 \GGt)} ~ g_0^2 ~\< \mc{O} 
    \widehat{\GGt}(x) \> - counterterms - \< \mc{O} \overline{X}^S(x) \>
    \label{eq:XS}
 \ee
where $Z_A^{(S)}$ and $Z_{(g_0^2 \GGt)}$ are mixing coefficients, $\widehat{\GGt}(x)$ is a lattice version of the corresponding classical operator and the counterterms make separately finite the operators $\Delta_{\mu} \hat{A}_{\mu}^S(x)$ and $g_0^2 ~ \widehat{\GGt}(x)$. A complete one-loop analysis of the renormalization constants appearing in Eqs.~(\ref{eq:Xa}) and (\ref{eq:XS}) for both the Wilson and Clover actions will be reported elsewhere~\cite{Diego}. The main difference between the singlet and non-singlet channels is that beyond one loop the divergence of the axial current $Z_A^{(S)} \Delta_{\mu} \hat{A}_{\mu}^S(x)$ and the anomaly $Z_{(g_0^2 \GGt)} ~ g_0^2 ~ \widehat{\GGt}(x)$ are not yet finite operators, and a logarithmically divergent mixing, represented by the counterterms in Eq. (\ref{eq:XS}), is required in order to end up with finite operators (see, e.g., Ref.~\cite{Testa}). Denoting such operators by $\Delta_{\mu} \hat{A}_{\mu}^{(S, ren.)}(x)$ and $g^2 \widehat{\GGt}^{(ren.)}(x)$, respectively, one gets
 \be
    & - & \Delta_{\mu} \< \mc{O} \left[ \hat{A}_{\mu}^{(S, ren.)}(x) \right]
    \> + \< \mc{O} \left[ Z_P^{(S)} \hat{\bar{\psi}}(x) \left\{ M^{(S)}, 
    \lambda_0 \right\} \gamma_5 ~ \hat{\psi}(x) \right] \> 
    \nonumber \\
    & + & 2 N_f {g^2 \over 32 \pi^2} \< \mc{O} \widehat{\GGt}^{(ren.)}(x) \> 
    = \< {\delta[\mc{O}] \over \delta(i\alpha^0(x))} \> + O(a)
    \label{eq:preWIS}
 \ee
where $M^{(S)} \equiv [M_0 - \overline{M}^{(S)}] / Z_P^{(S)}$ with $Z_P^{(S)}$ being the renormalization constant of the singlet pseudoscalar density. Therefore, in analogy with the non-singlet case one can identify the continuum limit of the operators $\hat{A}_{\mu}^{(S, ren.)}(x)$, $Z_P^{(S)} \hat{\bar{\psi}}(x) \left\{ M^{(S)}, \lambda_0 \right\} \gamma_5 ~ \hat{\psi}(x)$ and $\widehat{\GGt}^{(ren.)}(x)$ with the renormalized axial current $A_{\mu}^S(x)$, $\bar{\psi}(x) \left\{ M^{(S)}, \lambda_0 \right\} \gamma_5 ~\psi(x)$ and gluon anomaly $\GGt(x)$, respectively. In this way the singlet continuum $WI$ is recovered with a mass $M^{(S)}$. Although one has $Z_P^{(S)} \neq Z_P^{(NS)}$ and $\overline{M}^{(S)} \neq \overline{M}^{(NS)}$, it has been shown~\cite{Testa} that $[M_0 - \overline{M}^{(S)}] / Z_P^{(S)} = [M_0 - \overline{M}^{(NS)}] / Z_P^{(NS)}$, which means $M^{(S)} = M^{(NS)} = M$. In this way the continuum $WI$'s are preserved. In case of $SU(3)$-symmetry breaking the equivalence between $M^{(S)}$ and $M^{(NS)}$ is {\em essential} for the applicability of our strategy leading to Eq.~(\ref{eq:result}).

\section{Conclusions \label{section:conclusions}}

\indent In this paper we have presented a strategy for evaluating the neutron $EDM$ on the lattice induced by the strong $CP$ violating term of the $QCD$ Lagrangian. Our strategy is based on the definition of the neutron $EDM$ given by Eq.~(\ref{eq:EDM}), which involves the insertion of the topological charge in the presence of the charge density operator $J_0$. In case of three flavors with non-degenerate masses we have presented a complete diagrammatic analysis showing how the axial chiral Ward identities can be used to replace the topological charge operator with the flavor-singlet pseudoscalar density $P_S$. Our final result (\ref{eq:result}) is characterized only by disconnected diagrams, where the disconnected part can be either the single insertion of $P_S$ or the separate insertions of both $P_S$ and $J_0$. In this way we confirm the final outcome of Ref.~\cite{Aoki} that connected insertions of the pseudoscalar density do not contribute to the neutron $EDM$. Finally we have discussed the applicability of our strategy to the case of lattice formulations that break explicitly chiral symmetry, like the Wilson and Clover actions.

\section*{Acknowledgments} The authors gratefully acknowledge M. Testa for many useful discussions. This work is partially supported by the European Network ``Hadron Phenomenology and Lattice $QCD$", HPRN-CT-2000-00145.

\end{document}